# TESS Unveils the Full Phase Curve of WASP-19b

Mohammad Eftekhar[1] and Pouyan Adibi[2]
[1] University of Hormozgan Department of Physics, 3995 Bandar Abbas, Iran; m.eftekhar@znu.ac.ir
[2] University of Hormozgan Department of Mechanical Engineering, 3995 Bandar Abbas, Iran

## Abstract

We present the detection and analysis of the full-orbit phase curve and secondary eclipse of the short-period transiting hot Jupiter system WASP-19b with a single joint fit to photometric data and resolve parameter degeneracy. We analyze data taken by the Transiting Exoplanet Survey Satellite (TESS) during sectors 9 and 36. We model the data with our five-component model: primary transit, secondary eclipse, ellipsoidal variations, thermal emission, and reflected light, which are jointly fit to extract the information from all parameters simultaneously. The amplitude of Doppler beaming was also estimated to be ∼3 parts-per-million (ppm), but given the precision of the photometric data, we found that it was negligible and excluded it from the total phase curve model. We confidently report the secondary eclipse depth of $494^{+59}_{-48}$ ppm, the most accurate eclipse depth determined so far for WASP-19b, after cleaning the data from the instrumental systematic noise. According to the TESS bandpass, the day and nightside temperatures of WASP-19b are $2245^{+19}_{-20} K$ and $1095^{+20}_{-21}$, respectively. In addition, we find that the region of maximum brightness is well aligned with the substellar point, implying that there is an inefficient heat distribution from the dayside to the nightside. Our derived $A_g = 0.11^{+0.03}_{-0.03}$, suggesting that WASP-19b's geometric albedo is greater than the geometric albedos of most other hot Jupiters. Finally, the ellipsoidal variation signal amplitude we calculated agrees with theoretical expectations. Our comprehensive model with the approach of Markov Chain Monte Carlo shows the remedy of the degeneracy parameter in photometric data.

*Unified Astronomy Thesaurus concepts:* Exoplanet astronomy (486); Exoplanet systems (484)

## 1. Introduction

Since 2018 August, the Transiting Exoplanet Survey Satellite (TESS; Ricker et al. 2015) has been monitoring the southern and northern ecliptic hemispheres for planets orbiting bright stars. With a wavelength region of (0.6–0.95 μm), TESS can measure planetary light as a function of longitude, including thermally emitted and reflected light. The high-sensitivity of the TESS data has made it possible to study variability across a star–planet system's orbital phase.

The detection of secondary eclipse and characterization of the phase curve of WASP-121b (Bourrier et al. 2020; Daylan et al. 2021; Eftekhar 2022a, 2022b), WASP-18 (Shporer et al. 2019), KELT-1b (Beatty et al. 2020; von Essen et al. 2021; Eftekhar & Abedini 2022), and WASP-100b (Jansen & Kipping 2020), are of particular interest to this study. TESS has measured the phase curves of all of these ultrahot Jupiters, which have dayside temperatures of about ∼2200 K and are predicted to be tidally locked due to their close proximity to the host star.

Another ultrahot Jupiter that has been observed by TESS is WASP-19b. The bright host G-dwarf star ($V = 12.3$, $T_{eff} = 5568 \pm 71$ K; Torres et al. 2012; Mancini et al. 2013), short orbital period ∼0.788 days, and inflated radius ($a/R_s = 3.630$, $R_p = 1.41 R_J$), make it an attractive target for studying its atmosphere and secondary eclipse using various methods. Moreover, WASP-19b is also expected to have atmospheric characteristics detectable in the average planetary flux (Showman & Guillot 2002) due to it being tidally locked to its host star (Mazeh 2008). Several studies have determined WASP-19b's primary transit at optical and infrared wavelengths (e.g., Bean et al. 2013; Huitson et al. 2013; Mancini et al. 2013; Sedaghati et al. 2015, 2017; Espinoza et al. 2019; Wong et al. 2020a, 2020b). By using near-infrared, optical photometry, and spectroscopic data using the Hubble Space Telescope and broadband observations of four Spizter/IRAC bands, secondary eclipses were measured (Anderson et al. 2010; Burton et al. 2012; Bean et al. 2013; Lendl et al. 2013; Mancini et al. 2013; Zhou et al. 2014; Wong et al. 2020a, 2020b). According to an analysis of the thermal emission, the day and nightside temperatures of WASP-19b were measured to be $2240 \pm 40$ K and $1090^{+190}_{-250}$ K (Wong et al. 2020a), respectively, which is to be expected given the planet's proximity to its host G-type star and is comparable to other hot Jupiters with high irradiation. Based on the optical phase curve analysis by Wong et al. (2020a), the geometric albedo was calculated to be $0.16 \pm 0.040$. The reflection component is calculated with other parameters in our study to highlight the correlations between all of the constrained parameters.

The main objective of our study is to use our full phase curve joint model to learn more about the thermal emission and atmosphere structure of WASP-19b. For this purpose, we model the primary transit (assuming transit to be at the zero orbital phase), secondary eclipse, and three main phase curve components, including the tidal ellipsoidal distortion, reflection, and thermal emission. We also estimated Doppler beaming and determined that, given the precision of the TESS data, it is not significant. Using our comprehensive joint model, which allows us to extract information from all constrained parameters simultaneously, we can determine the uncertainties and correlations between all constrained parameters. The current analysis of the visible light using our joint model adds to the previous measurements mentioned in this section and comparisons of our results with other studies such as Wong et al. (2020a, 2020b).

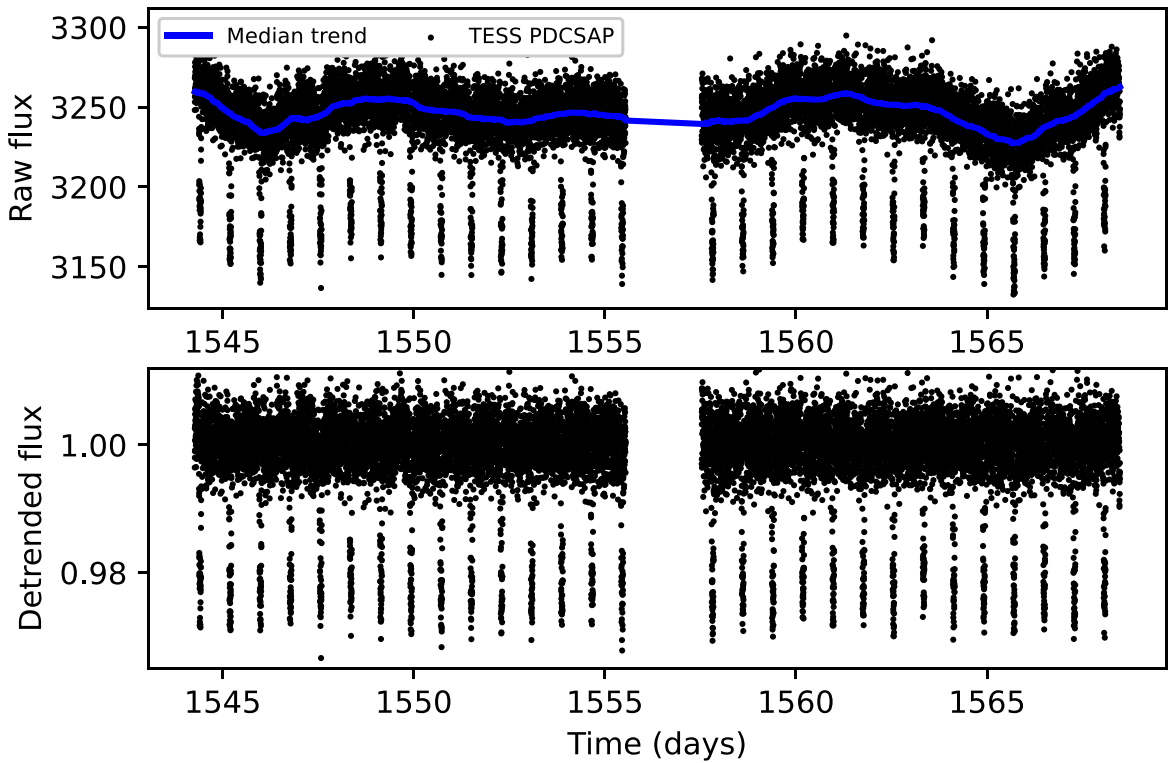

**Figure 1.** (Top) TESS light curve (PDCSAP flux) of WASP-19. The PDCSAP photometry is shown with black dots, and the solid blue line shows the trend generated by applying a detrending filter determined by wotan. Normalized PDCSAP light curve after median detrending (bottom).

The Paper is organized as follows: In Section 2, we describe the observations and data reduction methods. In Section 3, we explain the model for transit and secondary eclipse, and in Section 4, we present four main components that were utilized to characterize the phase curve of WASP-19b in detail. We detail the fitting procedure we used to get our results in Section 5. In Section 6, we present our physical parameters derived from TESS observations, and in Section 7, we explain them with a brief explanation.

## 2. Observations

WASP-19 (TIC 35516889; TOI 655) was observed by the TESS mission in sector 9, camera 2 (between 2019 February 28 and March 26), included in the list of preselected target stars. The downlinked data have a cadence of 2 minutes using an $11 \times 11$ pixel subarray centered on the target. To double the amount of data, we also include additional sector 36 data sets that are freely available (see Section 6). For the results presented in this work, we employed presearch data conditioning (PDC) light curves because they are cleaner than simple aperture photometry light curves (SAP) and exhibit considerably less scatter and short-timescale flux variations (Smith et al. 2012; Stumpe et al. 2014).

Even though the PDCSAP light curve's dominant systematics were corrected by default, we corrected it for further remaining systematics. To preserve variability within the planetary period, we flattened the PDCSAP light curve using the median detrending technique with a window length of one orbital period. The regression was performed using the Python package wotan (Hippke et al. 2019), as shown in Figure 1.

We applied phase folding to WASP-19b's orbital period after detrending and binned every 70 data points; the reprocessed data were utilized in our further analysis. The reprocessed data are shown in Figure 2.

## 3. Transit and Secondary Eclipse

We model both transit and secondary eclipses by utilizing the Mandel & Agol (2002) formalism implemented in the publicly available package batman (Kreidberg 2015). We set uninformative uniform priors on both midtransit time, $T_0$, and the orbital period, $P$, centered on the value reported in Wong et al. (2020a) to find an updated ephemeris and orbital period. We fit for the planet–star radius ratio, $p = R_p/R_s$, scaled orbital semimajor axis, $a/R_s$, and the relative brightness of the planets dayside hemisphere, $f_p$. We adopt a quadratic limb-darkening law for the transit model and fit the coefficients. In the fits shown here, we fix $e$ at zero, since photometric only data weakly constrains the orbital eccentricity. See Section 6 for further discussion on orbital eccentricity. Table 1 summarizes the uniform prior settings we adopted for this study.

## 4. Phase Curve

In addition to the transit and eclipse, visible light shows more variation along the orbital phase. While the depth of transit is sensitive to the planet–star radius ratio, the eclipse depth is related to the geometric albedo and the planets thermal emission. Also, modulations in visible light along the orbital phase are generated by longitudinal variations of the planets' brightness as well as gravitational interaction between the planet and star. The shape of the observed phase curve is a superposition of the effects of four main processes, which are described briefly below. Thermal emission, reflected light,

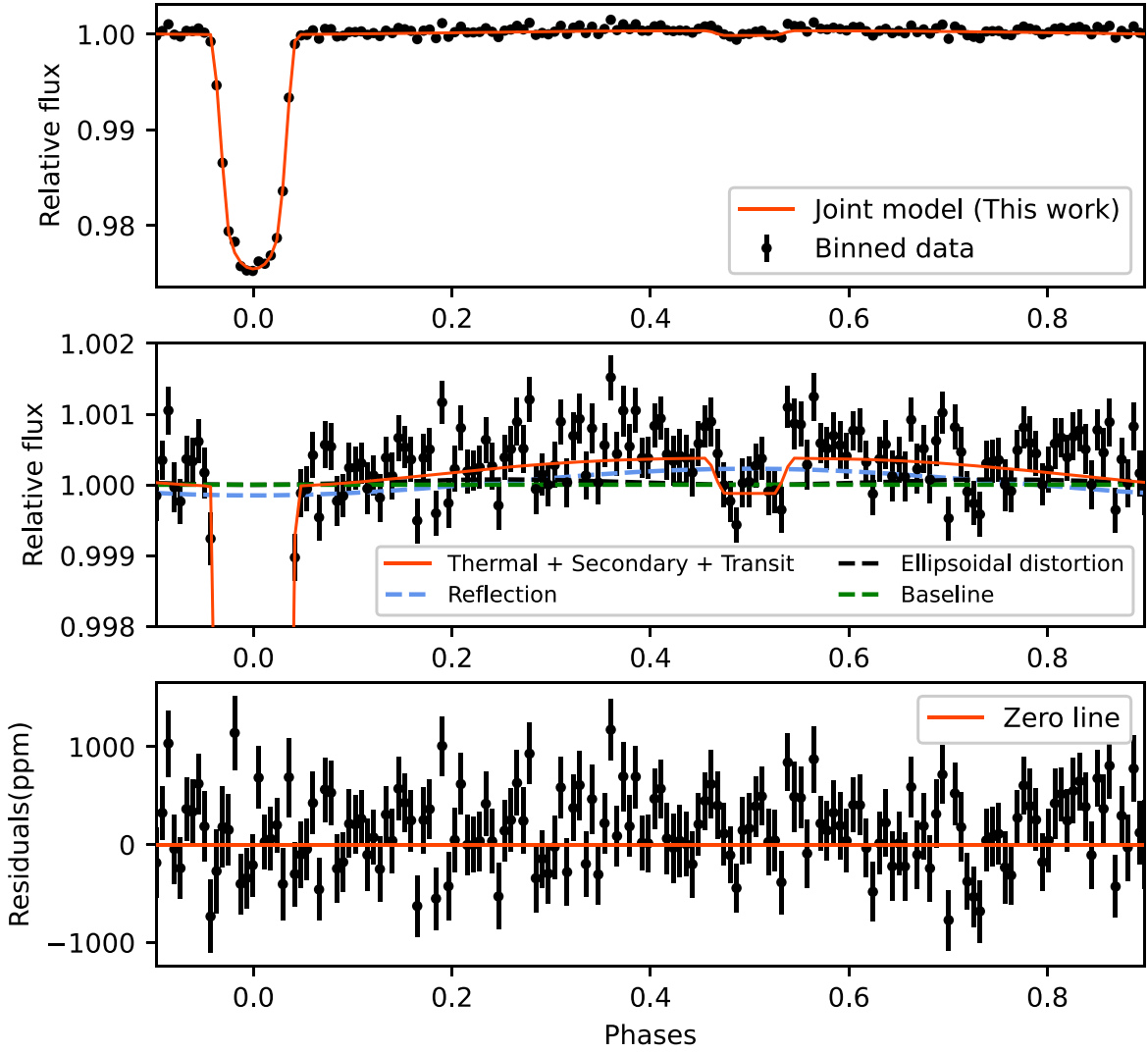

**Figure 2.** (Top) Our reprocessed data of WASP-19 (black dots) compared to our best-fitted full phase curve model (red curve). (Middle) Zoom of the secondary eclipse and phase curve variations with the reflection modulation (dashed blue curve), ellipsoidal distortion (dashed black curve), and baseline (dotted green line). (Bottom) The best-fitted model's corresponding residuals.

**Table 1**
Free Parameters, Uniform Priors' Range, and the Best-fitted Values

| Parameter | Prior | Value |
| --- | --- | --- |
| Orbital period; $P$ (days) | [0.1, 1.5] | $0.7888^{+0.0009}_{-0.0010}$ |
| Transit epoch; $T_0(BJD_{TDB})$ | [554, 556] | $555.45470^{+0.00020}_{-0.00021}$ |
| Planet–star radii ratio; $R_p/R_s$ | [0, 1] | $0.1513^{+0.0011}_{-0.0010}$ |
| Scaled semimajor axis; $a/R_s$ | [0, 6] | $3.6119^{+0.0008}_{-0.0009}$ |
| Orbital inclination $i$ (deg) | [0, 90] | $79.5^{+0.2}_{-0.2}$ |
| Limb-darkening coefficient; $u_1$ | [0, 1] | $0.268^{+0.068}_{-0.083}$ |
| Limb-darkening coefficient; $u_2$ | [0, 1] | $0.327^{+0.086}_{-0.073}$ |
| Radiative to advective timescales ratio; $\xi$ | [−10, 10] | $0.024^{+0.094}_{-0.077}$ |
| Nightside temperature; $T_N$ (K) | [0, 5000] | $1095^{+20}_{-21}$ |
| Day–night temperature difference; $\Delta T_{DN}$ (K) | [0, 2500] | $1101^{+19}_{-20}$ |
| Additive baseline | [−0.1, 0.1] | $-0.000046^{+8 \times 10^{-6}}_{-0.8 \times 10^{-6}}$ |
| Secondary eclipse depth (ppm) | [0, 800] | $494^{+59}_{-48}$ |
| Amplitude of the reflection; $A_{\text{ref}}$ (ppm) | [0, 500] | $199^{+44}_{-37}$ |
| Reflection phase shift; $\Delta_P$ | [−0.5, 0.5] | $0.0031^{+0.0056}_{-0.0066}$ |
| Amplitude of ellipsoidal variations; $A_{\text{ellip}}$ (ppm) | [0, 200] | $35^{+9}_{-10}$ |

Doppler beaming, and ellipsoidal variation are characteristics of these variations.

### 4.1. Thermal Emission

Although thermal emission predominates in the near-infrared (Knutson et al. 2012), it can also be detectable at visible wavelengths for highly irradiated exoplanets (Snellen et al. 2009). Due to tidal locking and closeness to the host star, WASP-19b amplifies the amplitude of the system's phase effects. Because of inefficient thermal redistribution, WASP-19b has a strong contrast between the dayside and nightside brightness temperatures (Wong et al. 2020a). As a result, WASP-19b is likely to have a zone (hotspot) with the maximum temperature and higher thermal flux compared to the rest of the planet.

To model the thermal emission component of WASP-19b, we employed a semiphysical model based on Zhang & Showman (2017), which was implemented in spiderman (Louden & Kreidberg 2018). Three parameters are utilized in this model to characterize the main features of the thermal light curve. The first one is the ratio of radiative to advection timescale, $\xi$, which controls the thermal phase shift. The longitudinal shift of the hotspot increases as $\xi$ rises. If $\xi$ rises, the nightside temperature also rises while the dayside temperature falls, resulting in a reduction of the temperature gap between day and nightside. The second one is the nightside temperature, $T_N$, which is the temperature on the planet's nightside. Finally, the temperature difference between day and night is shown by $\Delta T_{DN}$. For the host star, we employed the Phoenix model spectrum (Husser et al. 2013) and the TESS bandpass temperatures were calculated using `spiderman`.

### 4.2. Reflection

The reflected light phase curve is a portion of starlight that is reflected by the planet from its atmosphere or surface due to the geometric configuration of the planet–star in the bandpass of the observations. The orbital phase modulation of reflected light is expected to be sinusoidal and shows a maximum and minimum at superior conjunction and inferior conjunction, respectively (see Figure 2, dashed blue curve). The difference in reflectivity (albedo) determines the amplitude of the reflection (Shporer 2017). The following equation is a simple description of reflection phase modulation:

$$\text{Reflection} = A_{ref}(1 + \cos(2\pi(\phi + \Delta_P/P) + \pi)), \quad (1)$$

where $A_{\text{ref}}$ is the amplitude reflection, which is determined by the albedo, $\phi$ is the orbital phase, $P$ is the orbital period, and $\Delta_P$ is the phase shift. The geometric albedo of a planet, $A_g$, is defined as the ratio of its reflectivity at a zero phase angle to that of a Lambertian disk, and it can be calculated using the equations given below (Rodler et al. 2010):

$$A_g = A_{ref}(a/R_p)^2; \quad (2)$$

$a$ is the semimajor axis, and $R_p$ is the radius of the planet.

### 4.3. Doppler Beaming

The observed flux of a celestial object along our line of sight is affected by the relative radial velocity between the observer and the object due to the Doppler beaming effect. The orbital phase modulation of the Doppler beaming has a sinusoidal form with a maximum during the quadrature (0.25) phases, where the object is moving toward the observer, and at the quadrature (0.75) phases, where it is moving away. The amplitude of the beaming component, $A_{\text{beam}}$, can be computed in a simple way using the equations given below (Shporer 2017):

$$A_{\text{beam}} = 0.0028 \alpha_{\text{beam}} \left(\frac{P}{\text{day}}\right)^{-1/3}$$
$$\times \left(\frac{M_1 + M_2}{M_\odot}\right)^{-2/3} \left(\frac{M_2 \sin i}{M_\odot}\right), \quad (3)$$

where

$$\alpha_{\text{beam}} = \int \frac{1}{4} \frac{xe^x}{e^x - 1} dx, \quad x \equiv \frac{hc}{kT_{\text{eff}} \lambda}. \quad (4)$$

Here, the masses of the host star, planet, and Sun, are $M_1$, $M_2$, and $M_\odot$, respectively. The orbital inclination angle is $i$, $T_{\text{eff}}$ is the stellar effective temperature, $\lambda$ is the observed wavelength, and $h$ and $k$ are the Planck and Boltzmann constant, respectively. This integral should be taken in the TESS passband.

We found the amplitude of Doppler beaming to be $\sim$3 ppm using Equation (3), which is less than the precision achieved by TESS (even for a star as bright as WASP-19), so we decided to leave it out of our total phase curve model.

### 4.4. Ellipsoidal Variations

The tidal force that the planet induces on the star when the planet and star are close to each other is strong enough to tidally distort the star in a way that produces photometric orbital modulations. The distorted star deforms from spherical to ellipsoidal, a phenomenon known as ellipsoidal variations. We consider the formulation presented in Shporer (2017) for approximating the amplitude of the ellipsoidal photometric modulation, $A_{\text{ellip}}$;

$$A_{\text{ellip}} \simeq 13 \alpha_{\text{ellip}} \sin i \times \left(\frac{R_1}{R_\odot}\right)^3 \left(\frac{M_1}{M_\odot}\right)^{-2}$$
$$\times \left(\frac{P}{\text{day}}\right)^{-2} \left(\frac{M_2 \sin i}{M_J}\right) [ppm], \quad (5)$$

where

$$\alpha_{\text{ellip}} = 0.15 \frac{(15 + u)(1 + g)}{(3 - u)}. \quad (6)$$

Here, $u$ is the linear limb-darkening coefficient, and $g$ is the gravity-darkening coefficient. We obtain these value of $u$ and $g$ by linearly interpolating between metallicity, effective temperature $T_{\text{eff}}$, and $\log g$ from tabulation reported by Claret (2017). According to Equation (5), we found the amplitude of ellipsoidal variation is $\sim$32 ppm. The ellipsoidal variation can be described as:

$$\text{Ellipsoidial} = A_{\text{ellip}}(1 + \cos(4\pi\phi - \pi)). \quad (7)$$

## 5. Fitting Procedure

The comprehensive joint model in this work includes primary transit, secondary eclipse, and phase curves that consist of thermal emission, reflection, Doppler beaming, and ellipsoidal variations. A constant baseline was also added to compensate for any normalization bias. We simultaneously fit for all the mentioned components to extract information about all parameters from the data sets. It also allows us to understand the cases with degeneracies between parameters and the uncertainty and correlations between all of the constrained parameters. The problem of degeneracies between parameters is one of the most difficult aspects of simultaneous fitting. To overcome this problem, we have used the affine-invariant Markov Chain Monte Carlo (MCMC; Goodman & Weare 2010; Foreman-Mackey et al. 2013), using the `emcee` package (Foreman-Mackey et al. 2013).

We used an uninformative wide uniform prior to fit all parameters, allowing us to obtain their best estimation. The prior distributions that were chosen are listed in Table 1. We ran 200 walkers for 2000 steps, with a burn-in phase of the first 50% of each chain before obtaining posterior distributions. We plotted the chains and checked visually to ensure convergence. All of this process was repeated to ensure that the results were well within the 0.1$\sigma$. The median and standard deviation of posterior distributions were reported at $1\sigma$, which contains 68% of the posterior distribution.

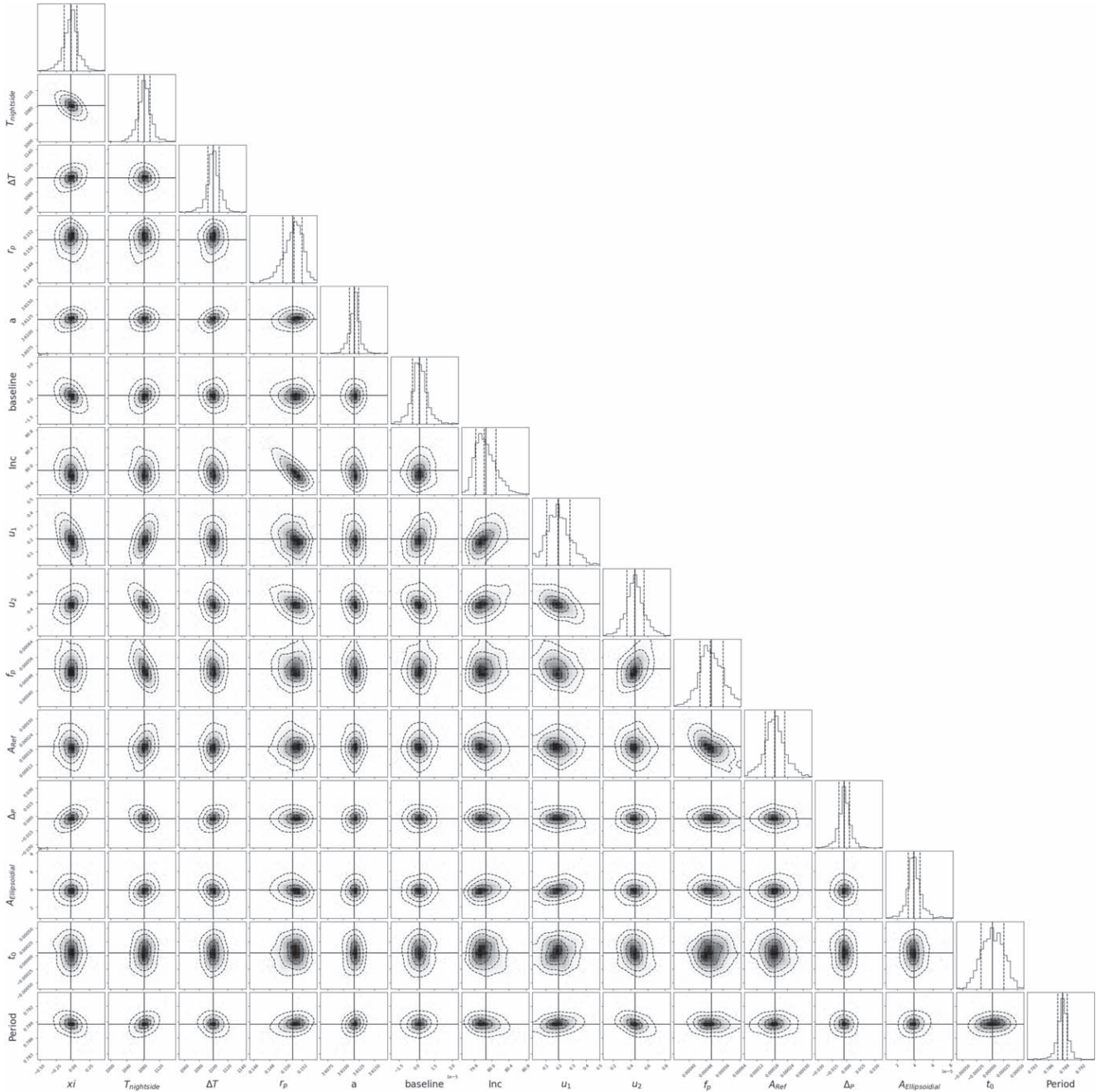

**Figure 3.** Retrieved posterior distributions by fitting our joint model to the phase curve of the WASP-19b. The black points indicate the best-fit values, and the colors of the contours highlight the $1\sigma$ and $2\sigma$ simultaneous 2D confidence regions that contain, respectively, 39.3% and 86.5% of the samples. The solid black line corresponds to the median values, while dashed black lines show the $1\sigma$ highest density intervals.

## 6. Result

Table 1 lists the median and $1\sigma$ uncertainties derived from the posterior distributions. The reprocessed data as well as the best-fitted model of the full phase curve are shown in Figure 2, and the corner plot for our retrieved posterior distributions from the joint model fit is shown in Figure 3. The best-fitted model's reduced chi-squared, $\chi^2$ percent, is 1.21, indicating a good fit to the TESS photometry.

We found that our results are mostly consistent with other reported values in the literature (Wong et al. 2020a, 2020b). The planetary radius $R_p/R_s$ and secondary eclipse depth are calculated to be $0.1513^{+0.0011}_{-0.0010}$ and $494^{+59}_{-48}$ ppm, respectively. Our measurement of the eclipse depth is the most precise to date and also it is well within $1\sigma$ of the value presented in Wong et al. (2020a, 2020b). We present an updated midtransit time, $T_0$, and the orbital period, $P$, which is consistent with the results of Wong et al. (2020a, 2020b), as well as studies that fitted ephemerides, (Hellier et al. 2011; Lanza 2020; Petrucci et al. 2020). Our MCMC yielded a quadratic limb-darkening coefficient of $u_1 = 0.268^{+0.068}_{-0.083}$ against 0.379 and $u_2 = 0.327^{+0.086}_{-0.073}$ against 0.205 reported in Claret (2017),

which is significantly different as well as consistent well within $1\sigma$ of the value presented in Wong et al. (2020b).

Using Equation (2), we estimate a significant dayside geometric albedo of $0.11^{+0.03}_{-0.03}$, which is consistent with the results of Wong et al. (2020a, 2020b).

We obtained the ratio of the radiative versus advection timescale of atmospheric height $\xi = 0.024^{+0.094}_{-0.077}$, which is statistically consistent with zero. This value implies that WASP-19b's atmosphere has inefficient heat redistribution from day to nightside, which is consistent with results in the literature (Wong et al. 2020a, 2020b) and theoretical models (Perez-Becker & Showman 2013; Komacek et al. 2017). Due to poor thermal redistribution (advection does not redistribute heat across longitudes; Zhang & Showman 2017), WASP-19b's maximum temperature zone is located close to the substellar point and results in substantial differences in the day and nightside temperatures ($1101^{+17}_{-19}K$). We also measured the temperature of the day and nightside of $2245^{+19}_{-20}K$ and $1095^{+20}_{-21}$, respectively, which is in agreement with the temperatures published in Wong et al. (2020a, 2020b) and places WASP-19b in the ultrahot Jupiter class (Bell & Cowan 2018; Parmentier et al. 2018).

According to theoretical estimations, the amplitude of ellipsoidal variation and Doppler beaming is substantially lower than that of reflected light and thermal emission (see Figure 2). Based on Equation (5), we obtained an ellipsoidal variation amplitude of $35^{+9}_{-10}$, which is more in line with theoretical estimates of $\sim$32 ppm. This implies that the physical formalism in the equation sufficiently describes the photometric signal resulting from the host star's ellipsoidal variation. This value is $(1.8\sigma)$ greater than the value determined by Wong et al. (2020b). Because our theoretical estimation of the amplitude of Doppler beaming yields a value of $\sim$3 ppm, which is not significant given the precision of photometric data, we did not include it in our phase curve model.

One of the most valuable outcomes of this research is that it simultaneously fits for the primary transit, secondary eclipse, and all phase curve components. The robust detection of the full phase curve component corresponding to thermal emission, reflected light, and ellipsoidal variation, as well as the orbital parameters are shown in the middle panel of Figure 2, and the corresponding fit parameters are listed in Table 1.

In this study, we have set the orbital eccentricity of $e$ to zero, as in previous works (Wong et al. 2020a, 2020b). We also investigated what would happen if we allowed the eccentric $e$ and $\omega$ to vary freely, yielding weak eccentricity constraints: $e = 0.0022^{+0.0043}_{-0.0051}$ and $\omega = 23^{+26}_{-67}$ deg. In this case, estimated parameters consistent with the best-fit values are reported in Table 1. Due to the Lucy–Sweeney bias (Lucy & Sweeney 1971), a result of $e > 2.45\sigma_e$ is required to measure a nonzero eccentricity with 95% confidence; $\sigma_e$ is the standard deviation of the eccentricities (Eastman et al. 2013). For this reason, we can confidently rule out WASP-19b nonzero eccentricity.

We also tested a condition in which the planetary flux is totally reflecting, in addition to our full phase curve model. To model planetary reflection, we employed the Lambertian reflection model, which is implemented in `spiderman` and characterized by a geometric albedo of $A_g$. In this case, we calculated the geometric albedo of $0.243^{+0.048}_{-0.044}$, which is quite close to $A_g = \delta(a_p/R_p)^2 = 0.28^{+0.04}_{-0.04}$, which is the value derived from the TESS secondary eclipse depth ($\delta = 494^{+59}_{-48}$ ppm). The reduced $\chi^2$ in this case is 1.8.

## 7. Summary and Conclusions

We have presented here a study of the full orbital phase curve of the short-period transiting hot Jupiter system WASP-19, measured by TESS in sectors 9 and 36. We initially utilized the median detrending approach with a window length of one orbital period to smooth the TESS data. We binned every 70 data points after phase folding at the orbital period. In our following study, we used this reprocessed data. So we fitted our joint model, which consists of primary transit, secondary eclipse, and phase curves that incorporate the thermal emission, reflection, and ellipsoidal variations.

After eliminating systematic noise, we confidently report the secondary eclipse with a depth of $494^{+59}_{-48}$ ppm. Due to the combination of thermal emission and reflection in the TESS bandpass, WASP-19b has a significant secondary eclipse depth.

Our $\xi = 0.024^{+0.094}_{-0.077}$ measurement is statistically consistent with zero. This value implies that WASP-19b's atmosphere has inefficient heat redistribution from day to nightside, which is consistent with results in the literature (Wong et al. 2020a) and theoretical models (Perez-Becker & Showman 2013; Komacek et al. 2017). It also agrees with other highly irradiated hot Jupiters that have been shown in Wong et al. (2020b). As advection does not redistribute heat across longitudes, this result can be interpreted as WASP-19b's maximum temperature zone being located near the substellar point (close to the substellar point) due to inefficient thermal redistribution (Zhang & Showman 2017). We note that the positive value of $\xi$ corresponds to winds going from west to east and shifting the hotspot eastward of the substellar point. This parameter can also take a negative value. As the value of $\xi$ increases, heat is redistributed more effectively across longitudes by advection until the heat distribution gets uniform. Due to the inefficient thermal redistribution, WASP-19b's night and dayside temperatures are significantly different ($1101^{+19}_{-20}$). WASP-19b, with a measured dayside temperature of $2245^{+19}_{-20}K$, belongs to the ultrahot Jupiter class (Bell & Cowan 2018; Parmentier et al. 2018).

We estimate the geometric albedo of $0.11^{+0.03}_{-0.03}$ in the TESS passband, which is consistent with the results of Wong et al. (2020a, 2020b). This value indicates that WASP-19b's geometric albedo is relatively higher than other short-period hot Jupiter exoplanets, such as WASP-18b ($A_g < 0.048$ at $2\sigma$; Shporer et al. 2019), Qatar-2b ($A_g < 0.06$ at $2\sigma$; Dai et al. 2017), and WASP-12b ($A_g < 0.064$ at 97.5% confidence; Bell et al. 2017) and other hot Jupiters at the same wavelength, particularly irradiated hot Jupiters (Schwartz & Cowan 2015). Also, the bond albedo is high when the geometric albedo is high, since TESS's bandpass is near to the wavelength region where the host star is brightest (Shporer et al. 2019). The high albedo on a planet with a $1500 < T_d < 3000~K$ is consistent with a possible correlation between increasing dayside temperature and increasing geometric albedo (Wong et al. 2021).

According to theoretical estimations, the amplitude of ellipsoidal variation and Doppler beaming is significantly smaller than that of reflected light and thermal emission (see Figure 2). Because our theoretical estimate of Doppler beaming amplitude yields a value of $\sim$3 ppm, which is not significant given the precision of photometric data, we did not include it in our phase curve model. Finally, our best-fit joint model provided us with an estimate of ellipsoidal variation, with an amplitude of $35^{+9}_{-10}$ ppm. The amplitude of this value is consistent with theoretical expectations of ∼32 ppm, as shown in Section 4.4. It is also (1.7$\sigma$) higher than the value calculated by Wong et al. (2020b).

WASP-19b is highly irradiated because of its hot host star and short orbital period. Furthermore, its high dayside temperature, lack of statistically significant phase shift, and poor thermal distribution are similar to those of other highly irradiated gas giant planets, and a relatively high geometric albedo agrees with a possible correlation between increasing temperature and increasing geometric albedo. This study is the first attempt to analyze a full phase curve of WASP-19b with a single joint fit, which indicates that our comprehensive model and the approach of MCMC can be used to explore the full phase curves of transiting hot Jupiters with short periods with the remedy of the degeneracy parameter in photometric TESS data. According to an analysis of the WASP-19b phase curve modulations, TESS data are sensitive to photometric variations in systems with short periods and massive planets.

## 8. Software and Third Party Data Repository Citations

The AAS journals would like to encourage authors to change software and third party data repository references from the current standard of a footnote to a first class citation in the bibliography. As a bibliographic citation these important references will be more easily captured and credit will be given to the appropriate people.

The first step to making this happen is to have the data or software in a long-term repository that has made these items available via a persistent identifier like a digital object identifier (DOI). A list of repositories that satisfy this criteria plus each one's pros and cons are given at https://github.com/AASJournals/Tutorials/tree/master/Repositories.

In this work, we used data collected by the TESS mission. Funding for the TESS mission is provided by the NASA Science Mission Directorate. NASA's Science Mission Directorate is funding the TESS mission. The collected data from this mission are available at the Mikulski Archive for Space Telescopes (MAST). Special thanks to Mahmoudreza Oshagh, who helped with useful suggestions that greatly improved the paper, and fruitful discussions on the topics covered in this paper.

*Facility:* TESS

*Software:* BATMAN (Kreidberg 2015), emcee (Foreman-Mackey et al. 2013), SPIDERMAN (Louden & Kreidberg 2018).

## ORCID iDs

Mohammad Eftekhar  https://orcid.org/0000-0003-1596-0197

## References

Anderson, D. R., Gillon, M., Maxted, P. F. L., et al. 2010, A&A, 513, L3
Bean, J. L., Désert, J.-M., Seifahrt, A., et al. 2013, ApJ, 771, 108
Beatty, T. G., Wong, I., Fetherolf, T., et al. 2020, AJ, 160, 211
Bell, T. J., & Cowan, N. B. 2018, ApJL, 857, L20
Bell, T. J., Nikolov, N., Cowan, N. B., et al. 2017, ApJL, 847, L2
Bourrier, V., Ehrenreich, D., Lendl, M., et al. 2020, A&A, 635, A205
Burton, J. R., Watson, C. A., Littlefair, S. P., et al. 2012, ApJS, 201, 36
Claret, A. 2017, A&A, 600, A30
Dai, F., Winn, J. N., Yu, L., & Albrecht, S. 2017, AJ, 153, 40
Daylan, T., Günther, M. N., Mikal-Evans, T., et al. 2021, AJ, 161, 131
Eastman, J., Gaudi, B. S., & Agol, E. 2013, PASP, 125, 83
Eftekhar, M. 2022a, ARep, 66, 606
Eftekhar, M. 2022b, RMxAA, 58, 99
Eftekhar, M., & Abedini, Y. 2022, NewA, 96, 101837
Espinoza, N., Rackham, B. V., Jordán, A., et al. 2019, MNRAS, 482, 2065
Foreman-Mackey, D., Hogg, D. W., Lang, D., & Goodman, J. 2013, PASP, 125, 306
Goodman, J., & Weare, J. 2010, Communications in Applied Mathematics and Computational Science, 5, 65
Hellier, C., Anderson, D. R., Collier-Cameron, A., et al. 2011, ApJL, 730, L31
Hippke, M., David, T. J., Mulders, G. D., & Heller, R. 2019, AJ, 158, 143
Huitson, C. M., Sing, D. K., Pont, F., et al. 2013, MNRAS, 434, 3252
Husser, T. O., Wende-von Berg, S., Dreizler, S., et al. 2013, A&A, 553, A6
Jansen, T., & Kipping, D. 2020, MNRAS, 494, 4077
Knutson, H. A., Lewis, N., Fortney, J. J., et al. 2012, ApJ, 754, 22
Komacek, T. D., Showman, A. P., & Tan, X. 2017, ApJ, 835, 198
Kreidberg, L. 2015, PASP, 127, 1161
Lanza, A. F. 2020, MNRAS, 497, 3911
Lendl, M., Gillon, M., Queloz, D., et al. 2013, A&A, 552, A2
Louden, T., & Kreidberg, L. 2018, MNRAS, 477, 2613
Lucy, L. B., & Sweeney, M. A. 1971, AJ, 76, 544
Mancini, L., Ciceri, S., Chen, G., et al. 2013, MNRAS, 436, 2
Mandel, K., & Agol, E. 2002, ApJL, 580, L171
Mazeh, T. 2008, in EAS Publications Ser., Vol.29, Tidal Effects in Stars, Planets and Disks, ed. M. J. Goupil & J. P. Zahn (Les Ulis: EDP Sciences), 1
Parmentier, V., Line, M. R., Bean, J. L., et al. 2018, A&A, 617, A110
Perez-Becker, D., & Showman, A. P. 2013, ApJ, 776, 134
Petrucci, R., Jofré, E., Gómez Maqueo Chew, Y., et al. 2020, MNRAS, 491, 1243
Ricker, G. R., Winn, J. N., Vanderspek, R., et al. 2015, JATIS, 1, 014003
Rodler, F., Kürster, M., & Henning, T. 2010, A&A, 514, A23
Schwartz, J. C., & Cowan, N. B. 2015, MNRAS, 449, 4192
Sedaghati, E., Boffin, H. M. J., Csizmadia, S., et al. 2015, A&A, 576, L11
Sedaghati, E., Boffin, H. M. J., MacDonald, R. J., et al. 2017, Natur, 549, 238
Showman, A. P., & Guillot, T. 2002, A&A, 385, 166
Shporer, A. 2017, PASP, 129, 072001
Shporer, A., Wong, I., Huang, C. X., et al. 2019, AJ, 157, 178
Smith, J. C., Stumpe, M. C., Van Cleve, J. E., et al. 2012, PASP, 124, 1000
Snellen, I. A. G., de Mooij, E. J. W., & Albrecht, S. 2009, Natur, 459, 543
Stumpe, M. C., Smith, J. C., Catanzarite, J. H., et al. 2014, PASP, 126, 100
Torres, G., Fischer, D. A., Sozzetti, A., et al. 2012, ApJ, 757, 161
von Essen, C., Mallonn, M., Cowan, N. B., et al. 2021, A&A, 648, A71
Wong, I., Benneke, B., Shporer, A., et al. 2020a, AJ, 159, 104
Wong, I., Shporer, A., Daylan, T., et al. 2020b, AJ, 160, 155
Wong, I., Kitzmann, D., Shporer, A., et al. 2021, AJ, 162, 127
Zhang, X., & Showman, A. P. 2017, ApJ, 836, 73
Zhou, G., Bayliss, D. D. R., Kedziora-Chudczer, L., et al. 2014, MNRAS, 445, 2746